\documentclass{llncs}
\usepackage{etex}
\usepackage{rotating}
\usepackage{algorithmic}
\usepackage{float}
\usepackage{makecell}
\usepackage[all]{xy}
\usepackage{amssymb}
\usepackage{amsmath}
\usepackage{array}
\usepackage{booktabs}
\usepackage{hyperref}
\usepackage{amssymb}
\usepackage{amsmath}
\usepackage{tcolorbox}
\usepackage{url}
\usepackage{color, colortbl}
\usepackage{tabularx}
\usepackage{longtable}
\usepackage{rotating}
\usepackage{setspace}
\usepackage{xtab}
\usepackage{graphicx}
\usepackage{epsfig}
\usepackage{mathtools}
\usepackage{xcolor}
\usepackage{listings}% http://ctan. org/pkg/listings
\usepackage{blindtext}
\usepackage{verbatim}
\usepackage[linesnumbered,lined,boxed,commentsnumbered,ruled,vlined]{algorithm2e}

\usepackage{comment}
\usepackage{listings}% http://ctan.org/pkg/listings
\newtheorem{preDefinition}{{\it Definition}}%
{\end{preDefinition}}%

\newenvironment{keywords}{
       \list{}{\advance\topsep by0.35cm\relax\small
       \leftmargin=1cm
       \labelwidth=0.35cm
       \listparindent=0.35cm
       \itemindent\listparindent
       \rightmargin\leftmargin}\item[\hskip\labelsep
                                     \bfseries Keywords:]}
     {\endlist}
\title{Cryptanalysis of two recently proposed ultralightweight authentication protocol for IoT}

 \author{Masoumeh Safkhani\inst{1},  Nasour Bagheri\inst{2}}
\institute{ Computer Engineering Department, Shahid Rajaee Teacher Training University, Tehran, Iran, Postal code: 16788-15811, Tel/fax:+98-21-22970117, \email{Safkhani@srttu.edu}
\and Electrical Engineering Department, Shahid Rajaee Teacher Training University, Tehran, Iran, \email{NBagheri@srttu.edu}}

\authorrunning{M. Safkhani, Y. Bendavid, S.   Rostampour, N. Bagheri}

\begin{document}
\maketitle
 
 %_________________________________________________________________________________________________________________________
\begin{abstract}

By expanding the connection of objects to the Internet and their entry to human life, the issue of security and privacy has become important. In order to enhance security and privacy on the Internet, many security protocols have been developed. Unfortunately, the security analyzes that have been carried out on these protocols show that they are vulnerable to one or few attacks, which eliminates the use of these protocols. Therefore, the need for a security protocol on the Internet of Things (IoT) has not yet been resolved.

Recently, Khor and Sidorov cryptanalyzed the Wang \textit{et al.}  protocol and presented an improved version of it. In this paper, at first, we show that this protocol also does not have sufficient security and so it is not recommended to be used in any application. More precisely, we present a full secret disclosure attack against this protocol, which extracted the whole secrets of the protocol by two communication with the target tag. 

In addition, Sidorv \textit{et al.} recently proposed an ultralightweight mutual authentication RFID protocol for blockchain enabled supply chains, supported by formal and informal security proofs. However, we present a full secret disclosure attack against this protocol as well. 
\end{abstract}
%__________________________________________________________________________________________________________________________
%__________________________________________________________________________________________________________________________
{\small
}
%__________________________________________________________________________________________________________________________
\section{Introduction} 

Internet of Things (IoT) can be used to control all devices through an Internet connection. IoT covers a wide range of applications at large and small scale. One of the solutions for implementing the IoT is the use of Radio Frequency Identification (RFID) which allows the connection of various objects via the Internet.

Internet technology uses objects in a variety of contexts and, given the availability of security and privacy, has the ability to make things easy for humans.
One of the most fascinating things about the IoT is its ability to smart homes, smart cities, use it in wearable gadgets and also use it in cars. Wearable gadgets include a series of sensors and use special software to collect their users' data; these gadgets ultimately analyze aggregated data such as health, fitness, and the like, and report the results to the user. The IoT can also make cars, with the help of special sensors and of course the Internet, both increase their safety and the optimal performance of internal components and passengers' safety.

One of the main issues that has always been raised about the IoT is its security issue. Since this technology connects a lot of devices through the Internet, hacking them can have irreparable losses, such as losing sensitive personal and economic information. One reason for this is that either security protocols have not been used to secure IoT or that protocols have not been adequately secured. Therefore, a researchers' challenge in this field is to ensure users' privacy and the confidentially of their information by developing suitable software and hardware, to be used in Internet-based products. On the other hand, the use of security mechanisms is one of the main prerequisites for protecting the privacy and confidentiality of a variety of applications, and IoT is not an exception.

\subsection{Related Work and Motivation}\label{re} 

A factor that can help expand the IoT is to create confidence for users of this technology in terms of maintaining their privacy and security, because if there is no proper security in its infrastructure, damage to IoT-based equipment, the possibility of losing personal information, the loss of privacy, and even the disclosure of economic and other data, are highly likely, and this may lead to the inability to use it in critical applications.  Ronen and Shamir in \cite{ronen2017iot} pointed out that if security is not taken into account in the IoT-based infrastructure, the technology will threaten the future of the world as a nuclear bomb.

Reviewing the proposed mechanisms and designing new models that are compatible with IoT-based devices are also very important. Given that an IoT system can include many objects with limited resources, it requires special protocols to ensure that privacy and security are guaranteed. Therefore, with the further development of IoT, its security concerns are expected to receive more attention. So far, several security protocols have been proposed to ensure IoT security, e.g.~\cite{mukherjee2017end,taylor2017access,wang2017privacy,fan2017ulmap,serhrouchni2017lightweight}, however, most of them have failed in providing their security goals~\cite{rahman2016security,wang2017security,yincryptanalysis,aghili2018impersonation,sarvabhatla2014cryptanalysis,baghery2014game} and various attacks, such as the  protocol's secret values disclosure, DoS, traceability, impersonation and etc. were reported against them. The presentation of these attacks resulted in the development of the protocol's designing   knowledge and the protocol designers are designing their protocols in such a way as to be as safe as the published attacks so far. Unfortunately, there are still attacks against newly designed protocols, and this science has not yet matured.

Among different designing strategies of security protocols, attempts to design a secure ultralightweight protocol for constrained environment has a long (unsuccessful) history. Pioneer examples include \texttt{SASI}~\cite{sasi}, \texttt{RAPP}~\cite{Tian}, \texttt{SLAP}~\cite{SLAP}, \texttt{LMAP}~\cite{LMAP} and \texttt{R$^2$AP}~\cite{Zhuang0C14} and among the recent proposals is \texttt{SecLAP}~\cite{AGHILIFGCS}, and many other broken protocols that have been compromised by the later third parties analysis~\cite{Avoine,AhmadianIPL,Phan,GUMAP,cryptoeprint,safkhani2017passive,BagheriSPT14,Safkhani2018}.
All those protocols tried to provide enough security only using few lightweight operations such as bitwise operations, e.g. logical AND, OR, XOR and rotation. However, the mentioned analysis have shown that it is not easy to design a strong protocol using cryptographically-weak components.

In the line of designing ultralightweight protocols, in~\cite{tewari2017cryptanalysis}, Tewari and Gupta proposed a new ultralightweight authentication protocol for IoT and claimed that their protocol satisfies all security requirements. However, in \cite{safkhani2017passive}, an efficient passive secret disclosure attack is applied to this protocol. Moreover, in \cite{wang2017security}, Wang \textit{et al.} cryptanalyzed the Tewari and Gupta protocol and also proposed an improved version of it. This protocol later analysed by Khor and Sidorov~\cite{khor2018weakness}, where they also proposed an improved protocol following the same designing paradigm. In this paper, we consider the security of this improved protocol which has been proposed by Khor and Sidorov, and for simplicity, we call it \texttt{KSP} (stands for Khor and Sidorov protocol) and show that \texttt{KSP} is vulnerable to desynchronization attack and also against secret disclosure attack.

As a new emerging technology, blockchain is believed to provide higher data protection, reliability,  transparency, and lower management costs compared to a conventional centralized database. Hence, it could be a promising solution for large scale IoT systems.  Targeting those benefits Sidorov \textit{et al.} recently proposed an ultralightweight mutual authentication RFID protocol for blockchain-enabled supply chains~\cite{sidorov2019}. Although they have claimed security against various attacks, we present an efficient secret disclosure attack on it. For the sake of simplicity, we call this protocol \texttt{SOVNOKP}.

\subsection{Paper organization}
The rest of the paper is structured as follows: Section~\ref{pro} is dedicated to a brief review of \texttt{KSP} and \texttt{SOVNOKP} which are among recent proposals to provide security in the IoT environment. Our proposed secret disclosure attacks are applied to these protocols in Section~\ref{SOVNOKPattack}. Finally, the paper is concluded in Section ~\ref{conc}.

\section{ Description of \texttt{KSP} and \texttt{SOVNOKP}}\label{pro} 

To describe we use the notations that are represented in Table~\ref{notation}. These notations are also used through the other parts of the paper.

\begin {table} 
\caption {Notations used in the protocols' description}\label{notation} 
\begin{center} 
\begin{tabular}{ |l|p{6cm}| } 
\hline 
\multicolumn{2}{|c|}{Notations~~~~~ ~~~~~~~~~~~~~~~~~~~ ~~~~~ ~~~~~ ~~~~~ Description} \\ 
\hline 
$\mathcal{T}$ & An RFID tag\\ 
\hline
$\mathcal{R}$ & An RFID Reader\\ 
\hline
$\mathcal{A}$ & The Adversary\\ 
\hline
$IDS$ & The pseudonyms of the tag\\
\hline 
$ID$ & The identifier of the tag\\ 
\hline
$K$ & The tag's key\\ 
\hline 
$m,n,q$ and $r$ & The random numbers which are generated by the reader/tag\\
\hline 
$X_i$ & The $i^{th}$ bit of string  $X$\\
\hline 
$E(M,K)$ & The encryption function  which encrypts $M$ using $K$  as its key\\
\hline 
$wt(Y)$ & The Hamming weight of $Y$   which equals with the number of 1’s in $Y$ \\ 
\hline
$Rot(X,Y)$ & The left rotation of $X$ by amount of $Y$ mod $L$, where $L$ is the bit-length of $X$ \\
\hline 

$RRot(X,Y)$ & The right rotation of $X$ by amount of $Y$ mod $L$, where $L$ is the bit-length of $X$ \\
\hline 
%	$ X >>> Y=ROR(x,y)$ & The r
%	$ X >>> Y=ROR(x,y)$ & The right rotation of $X$by amount of $wt(Y)$\\ 
%$ L$ & The length of protocol parameters \\ 
$ \oplus $ & The bit wise exclusive-or operation\\ 
\hline
$h(.)$ & SHA-256 hash function\\ 
\hline	

\end{tabular} 
\end{center} 
\end {table}  
\subsection{\texttt{KSP}}
Tewari and Gupta recently~\cite{tewari2017cryptanalysis} proposed an ultra-lightweight authentication protocol for IoT environment. Soon after that, two analyzes were published on it~\cite{safkhani2017passive, wang2017security}. In one of them, i.e. ~\cite{ wang2017security}, Wang \textit{et al.} presented a secret disclosure attack that reveals the shared secret key between the server and the tag. To overcome this security flaw, they suggested the modifications on the protocol in such a way that it is expected to be secure against that attack. However, later Khor and Sidorov~\cite{khor2018weakness} have shown that Wang \textit{et al.}'s protocol is as insecure as its predecessor and also proposed a new improved protocol which we call it \texttt{KSP}. As it is shown in Fig. ~\ref{KSP}, following the notations represented in Table ~\ref{notation}, \texttt{KSP} runs as follows:

\begin{enumerate}

\item  The reader starts the authentication phase of the protocol by sending $Hello$ and a random number $r$ message to the tag.
\item The tag, when receives the message, generates a random number $q$ and computes $s=Rot(q\oplus r, wt(q))$, $T=IDS_{new}\oplus r\oplus q$ and $U=Rot(IDS_{old}\oplus IDS_{new}, wt(q))$ and sends ${s\|T\|U}$  to the reader.
\item Upon receipt of the message, the reader does as follows:

\begin{itemize}
\item It does an exhaustive search to find $wt(q)$  to find a record of its database that  satisfies the received $s$ and $T$, where the reader will be able to find $IDS_{old}$ also, given $U$, to authenticate the tag.
\item The reader generates two new random numbers $m$ and $n$ and computes $P=m \oplus n \oplus q$, $Q=Rot(n, wt(K_{new}))$ and $R=Rot(Rot(K \oplus m, wt(n)), wt(K\oplus m))$ and sends them to the tag.

\end{itemize}
\item Once the tag receives the message, it:
\begin{itemize}
\item Extracts $n$ and $m$ and verifies the messages integrity based on $R$ to authenticate the reader.
\item If the reader has been authenticated, the tag and the reader will update their new parameters
$IDS_{new}=Rot(IDS_{new}\oplus q, wt(n))\oplus Rot(q,wt(m))$ and $K_{new}=Rot(K_{new}\oplus q\oplus n, wt(m))\oplus Rot(m,wt(n))$.
\end{itemize}

\end{enumerate}

\begin{figure} 
\begin{center} 
\tiny{ 
\begin{tabular}{|p{4cm}|c|p{4cm}|} 
\hline \vspace{-5pt} 
& &\\ 
~~~~~~~~~~\textbf{Reader $R_j$} & &~~~~~~~~~~~~~~~~~~~~~~~~~~~~\textbf{Tag $T_r$}\\ 
$IDS^{old},K^{old}$,$IDS^{new},K^{new}$ & &$IDS^{old},K^{old}$,$IDS^{new},K^{new}$\\ 
\hline 
\hline \vspace{-4pt} 
& &\\ 
Generates $r$& $\xrightarrow{~~~Hello,r~~~}$&Generates $q$, Computes $s=Rot(q\oplus r, wt(q))$, $T=IDS_{new}\oplus r\oplus q$ and $U=Rot(IDS_{old}\oplus IDS_{new}, wt(q))$ \\ 
Extracts tag's records; Authenticates the tag,  Generates $m$ and $n$; Computes $P=m \oplus n \oplus q$, $Q=Rot(n, wt(K_{new}))$ and $R=Rot(Rot(K \oplus m, wt(n)), wt(K\oplus m))$; Updates parameters. & $\xleftarrow{~~~~s\|T\|U~~~~}$ & \\ 
& $\xrightarrow{~P\|Q\|R~}$& Extracts $m$ and $n$; Authenticates the reader;  Updates parameters. \\ 

\hline 
\hline 
$IDS^{old}=IDS_{new}$  && $IDS^{old}=IDS_{new}$\\ 
$K^{old}=K_{new}$  && $K^{old}=K_{new}$\\ 
$IDS_{new}=Rot(IDS_{new}\oplus q, wt(n))\oplus Rot(q,wt(m))$&&$IDS_{new}=Rot(IDS_{new}\oplus q, wt(n))\oplus Rot(q,wt(m))$\\
$K_{new}=Rot(K_{new}\oplus q\oplus n, wt(m))\oplus Rot(m,wt(n))$&&$K_{new}=Rot(K_{new}\oplus q\oplus n, wt(m))\oplus Rot(m,wt(n))$\\

\hline 
\end{tabular} 
}\end{center} 
\caption{Mutual authentication phase of  \texttt{KSP}~\cite{khor2018weakness}}\label{KSP} 

\end{figure}

\subsection{\texttt{SOVNOKP}}	
\begin{figure}
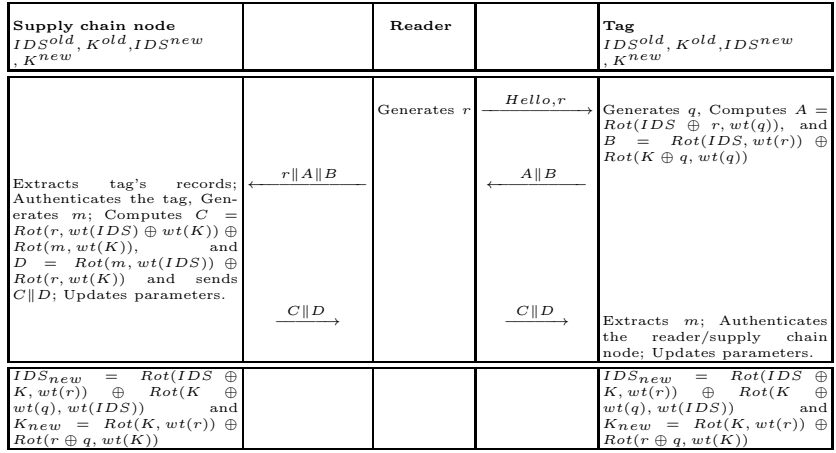
 
\begin{center} 
\tiny{ 
\begin{tabular}{|p{3cm}|c|c|c|p{3cm}|} 
\hline \vspace{-5pt} 
&&& &\\ 
\textbf{Supply chain node}&&\textbf{Reader } & &\textbf{Tag}\\ 
$IDS^{old},K^{old}$,$IDS^{new}$ &&& &$IDS^{old},K^{old}$,$IDS^{new}$\\ 
$,K^{new}$ &&& &$,K^{new}$\\ 
\hline 
\hline \vspace{-4pt} 
& &&&\\ 
&&Generates $r$& $\xrightarrow{~~~Hello,r~~~}$&Generates $q$, Computes $A=Rot(IDS\oplus r, wt(q))$, and $B=Rot(IDS, wt(r)) \oplus Rot(K\oplus q, wt(q))$ \\ 
Extracts tag's records; Authenticates the tag,  Generates $m$; Computes $C=Rot(r, wt(IDS)\oplus wt(K))\oplus Rot(m, wt(K))$, and $D=Rot(m, wt(IDS)) \oplus Rot(r, wt(K))$ and sends ${C\|D}$; Updates parameters. & $\xleftarrow{~~~~r\|A\|B~~~~}$ &&$\xleftarrow{~~~~A\|B~~~~}$ & \\ 
& $\xrightarrow{~C\|D~}$&	& $\xrightarrow{~C\|D~}$& Extracts $m$; Authenticates the reader/supply chain node;  Updates parameters. \\ 

\hline 
\hline 
$IDS_{new}=Rot(IDS\oplus K, wt(r))\oplus Rot(K\oplus wt(q),wt(IDS))$ and $K_{new}=Rot(K, wt(r))\oplus Rot(r\oplus q,wt(K))$&&&&$IDS_{new}=Rot(IDS\oplus K, wt(r))\oplus Rot(K\oplus wt(q),wt(IDS))$ and $K_{new}=Rot(K, wt(r))\oplus Rot(r\oplus q,wt(K))$\\

\hline 
\end{tabular} 
}\end{center} 
\caption{Mutual authentication phase of  \texttt{SOVNOKP}~\cite{sidorov2019}}\label{SOVNOKP}

\end{figure}

\texttt{SOVNOKP} protocol ~\cite{sidorov2019}follows the \texttt{KSP} protocol designing paradigm and it is based on almost similar components.  As it is shown in Fig. ~\ref{SOVNOKP}, following the notations represented in Table ~\ref{notation}, \texttt{SOVNOKP} runs as follows:
\begin{enumerate}

\item  The reader starts the authentication phase of the protocol by sending $Hello$ and a random number $r$ message to the tag.
\item The tag, when receives the message, generates a random number $q$ and computes $A=Rot(IDS\oplus r, wt(q))$, and $B=Rot(IDS, wt(r)) \oplus Rot(K\oplus r, wt(q))$ and sends ${A\|B}$  to the reader.
\item Upon receipt of the message, the reader forwards $r\|A\|B$ to the supply chain node.
\item Upon receipt of  $r\|A\|B$ , the reader does as follows:	
\begin{itemize}
\item It does an exhaustive search to find $wt(q)$  to find a record of its database that satisfies the received $A$, where the reader will be able to find $IDS$ and $K$ to verify the received $B$ to authenticate the tag.
\item Based on the $h(IDS\|K)$, \textit{it can check and track the product history together with permission level by reading the data from the blockchain. If the product has a correct history record in terms of ownership, timestamp, location, and product status, the supply chain node can authenticate the tag.}
\item Once the tag has been authorized, the supply chain node generates a random number $m$ and computes $C=Rot(r, wt(IDS)\oplus wt(K))\oplus Rot(m, wt(K))$, and $D=Rot(m, wt(IDS)) \oplus Rot(r, wt(K))$ and sends ${C\|D}$  to the reader.

\end{itemize}
\item The reader forwards the received ${C\|D}$  to the tag.
\item Once the tag receives the message, it:
\begin{itemize}
\item Extracts $m$  from $C$ and verifies the messages integrity based on $D$ to authenticate the reader/supply chain node.
\item If the reader/supply chain node has been authenticated and $wt(m)$ is even, the tag  will update its parameters as:
$IDS_{new}=Rot(IDS\oplus K, wt(r))\oplus Rot(K\oplus wt(q),wt(IDS))$ and $K_{new}=Rot(K, wt(r))\oplus Rot(r\oplus q,wt(K))$.
\end{itemize}

\end{enumerate}

\section{Security analysis of \texttt{KSP} and \texttt{SOVNOKP} }\label{SOVNOKPattack} 

Although the designers of \texttt{KSP} and \texttt{SOVNOKP} argued its security formally using GNY logic, however, it is easy to show that they are as vulnerable to different attacks as their predecessor rotation based protocols, e.g. see~\cite{SafkhaniBS18}.  

\subsection{Secret Disclosure Attack on \texttt{KSP}}\label{sdKSP}

To analyze \texttt{KSP}, we follow the same adversary model as them, when they analyzed Wang \textit{et al.}'s protocol~\cite[Sec.II, page 92]{khor2018weakness}, where \textit{``adversary is able to eavesdrop, intercept, block, and
modify messages sent during communication
between the reader and a tag.''}. Given the target tag, the attack procedure will be as follows:

%\begin{algorithm} 
%	% \hline\\ 
%	\KwData{$r,s, T,U,P,Q,R,r',s', T',U',P',Q',R'$} 
%	%\hline\\ 
%	\KwResult{Obtains $m$, $K$ } 
%	%\hline\\ 
%	$T\oplus T'=q\oplus r\oplus q'\oplus r'$\\
%	
%	$c=wt(q)$, $c'=wt(q')$\\ 
%	
%	$RRot(s,c) \oplus RRot(s',c')=T\oplus T'$ \\
%	$RRot(s,c)=q\oplus r$ ,$RRot(s',c')=q'\oplus r'$\\
%	$IDS_{new}=T\oplus RRot(s,c)$, $q=RRot(s,c) \oplus r$ \\
%	$IDS_{old}=IDS_{new}\oplus RRot(U, wt(q))$.\\
%	finds three offset values $c_1=wt(K_{new})$, $c_2=wt(n)+wt(K_{new}\oplus m)$ and $c_3=wt(n')+wt(K_{new}\oplus m')$\\
%	$P\oplus P'\oplus q\oplus q'=RRot(Q,c_1) \oplus RRot(Q',c_2) \oplus RRot(R,c_2)\oplus RRot(R',c_3)$.\\
%	$n=RRot(Q,c_1)=$,$n'=RRot(Q',c_2)$, $K\oplus m=RRot(R,c_2)=$ and $K\oplus m'=RRot(R',c_3)$ 
%	$m=P\oplus RRot(Q,c_1)\oplus q$ \\ $K=RRot(R,c_2)\oplus m $\\
%	%\hline\\ 
%	\caption{The algorithm of proposed secret disclosure attack against the \texttt{KSP} protocol }\label{al1} 
%\end{algorithm} 

\begin{enumerate}
\item  The adversary eavesdrops a session of the protocol between the target tag and a legitimate reader, where, the reader generates an arbitrary nonce $r$ and sends it to the tag, along with $Hello$.
\item The tag, when receives the message, generates a random number $q$ and computes $s=Rot(q\oplus r, wt(q))$, $T=IDS_{new}\oplus r\oplus q$ and $U=Rot(IDS_{old}\oplus IDS_{new}, wt(q))$ and sends ${s\|T\|U}$  to the reader, which is eavesdropped by the adversary.
\item Upon receipt of the message, the reader does as follows:
\begin{itemize}
\item It does an exhaustive search to find $wt(q)$  to find a record of its database that  satisfies the received $s$ and $T$, where the reader will be able to find $IDS_{old}$ also, given $U$, to authenticate the tag.
\item The reader generates two new random numbers $m$ and $n$ and computes $P=m \oplus n \oplus q$, $Q=Rot(n, wt(K_{new}))$ and $R=Rot(Rot(K \oplus m, wt(n)), wt(K\oplus m))$ and sends them to the tag, which is eavesdropped by the adversary.
\end{itemize}	

\item The adversary terminates the protocol.
\item  Given that tag has not received $P\|Q\|R$, it will not update its records, i.e. $IDS_{new}$, $K_{new}$, $IDS_{old}$ and  $K_{old}$. 
\item The adversary waits for another session between the target tag and a legitimate reader, where, the reader generates an arbitrary nonce $r'$ and sends it to the tag, along with $Hello$.
\item The tag, when receives the message, generates a random number $q'$ and computes $s'=Rot(q'\oplus r', wt(q'))$, $T'=IDS_{new}\oplus r'\oplus q'$ and $U'=Rot(IDS_{old}\oplus IDS_{new}, wt(q'))$ and sends ${s'\|T'\|U'}$  to the reader, which is eavesdropped by the adversary.
\item Upon receipt of the message, the reader does as follows:
\begin{itemize}
\item It does an exhaustive search to find $wt(q')$  to find a record of its database that  satisfies the received $s'$ and $T'$, where the reader will be able to find $IDS_{old}$ also, given $U'$, to authenticate the tag.
\item The reader generates two new random numbers $m'$ and $n'$ and computes $P'=m' \oplus n' \oplus q'$, $Q'=Rot(n', wt(K_{new}))$ and $R'=Rot(Rot(K \oplus m', wt(n')), wt(K\oplus m'))$ and sends them to the tag, which is eavesdropped by the adversary.
\end{itemize}	

\item Given that $T\oplus T'=q\oplus r\oplus q'\oplus r'$, the adversary finds two offset values $c$ and $c'$, respectively as $wt(q)$ and $wt(q')$, such that   and $RRot(s,c) \oplus RRot(s',c')=T\oplus T'$ and assigns $RRot(s,c)$ to $q\oplus r$ and $RRot(s',c')$ to $q'\oplus r'$.
\item The adversary now extracts  $IDS_{new}=T\oplus RRot(s,c)$, $q=RRot(s,c) \oplus r$ and $IDS_{old}=IDS_{new}\oplus RRot(U, wt(q))$.
\item  The adversary finds three offset values $c_1$, $c_2$ and $c_3$, respectively as $wt(K_{new})$, $wt(n)+wt(K_{new}\oplus m)$ and  $wt(n')+wt(K_{new}\oplus m')$, such that   and $P\oplus P'\oplus q\oplus q'=RRot(Q,c_1) \oplus RRot(Q',c_2) \oplus RRot(R,c_2)\oplus RRot(R',c_3)$.
\item The adversary assigns  $RRot(Q,c_1)$, $RRot(Q',c_2)$, $RRot(R,c_2)$ and $RRot(R',c_3)$ respectively to $n$, $n'$, $K\oplus m$ and $K\oplus m'$.
\item The adversary extracts $m=P\oplus RRot(Q,c_1)\oplus q$ and $K=RRot(R,c_2)\oplus m $.
\end{enumerate}	
Following above attack, the adversary extracts whole shared parameters between the tag and the reader with the complexity of eavesdropping two sessions of the protocol, blocking a message and doing polynomial computations.
It is clear that $T\oplus T=q\oplus r\oplus q'\oplus r'$. Now the adversary can also compute the updated values of $IDS_{new}$ and $K_{new}$ as $IDS_{new}=Rot(IDS_{new}\oplus q, wt(n))\oplus Rot(q,wt(m))$ and $K_{new}=Rot(K_{new}\oplus q\oplus n, wt(m))\oplus Rot(m,wt(n))$.
\begin{remark}:
It should be noted the adversary can use other metrics to filter possible wrong guesses, e.g., $c_1$, $c_2$ and $c_3$ should be respectively equal to  $wt(K_{new})$, $wt(n)+wt(K_{new}\oplus m)$ and  $wt(n')+wt(K_{new}\oplus m')$. 
\end{remark}
Given that the adversary has whole secret parameters, applying any other attack will be trivial, e.g., reader/tag impersonation, desynchronization and traceability attacks.

\subsection{Secret Disclosure Attack on \texttt{SOVNOKP}}\label{sdSOVNOKP}

To analyze \texttt{SOVNOKP}, we follow the same adversary model as them, when they analyzed Wang \textit{et al.}'s protocol~\cite[Sec.IV, page 9]{sidorov2019}, where \textit{`` adversary has the capability to initiate communication with the reader and a tag and it is able to eavesdrop, intercept, block, and
modify messages sent during communication
between the reader and a tag. Moreover, the number of unsuccessful attempts is limited to 5, otherwise, the reader sends KILL command to the tag ''}. The parameter length is considered to be 96 bits. Given the target tag, the attack procedure, which is almost similar to the attack against \texttt{KSP}, will be as follows:

\begin{enumerate}
\item  The adversary eavesdrops all transferred messages over a session of the protocol between the target tag and a legitimate reader, i.e., $r,A,B,C,D$ where, $A=Rot(IDS\oplus r, wt(q))$,  $B=Rot(IDS, wt(r)) \oplus Rot(K\oplus q, wt(q))$, $C=Rot(r, wt(IDS)\oplus wt(K))\oplus Rot(m, wt(K))$, and $D=Rot(m, wt(IDS)) \oplus Rot(r, wt(K))$.
\item The adversary blocks $C\|D$, sent from the reader to the tag. Hence, tag will not update its secrets, i.e., $IDS$ and $K$.
\item The adversary waits for another session between the the target tag and a legitimate reader, where, the reader generates an arbitrary nonce $r'$ and sends it to the tag, along with $Hello$.
\item The adversary again  eavesdrops all transferred messages over a session of the protocol between the target tag and a legitimate reader, i.e., $r',A',B',C',D'$ where, $A'=Rot(IDS\oplus r', wt(q'))$,  $B'=Rot(IDS, wt(r')) \oplus Rot(K\oplus q', wt(q'))$, $C'=Rot(r', wt(IDS)\oplus wt(K))\oplus Rot(m', wt(K))$, and $D'=Rot(m', wt(IDS)) \oplus Rot(r', wt(K))$.

\item Given that $RRot(A, wt(q))\oplus RRot(A', wt(q'))=IDS\oplus r\oplus IDS\oplus r'=r\oplus r'$ and $r$ and $r'$ are transferred over insecure channel, which is eavesdropped by the  adversary, the adversary extracts $IDS$ and $K$ as follows:
\begin{enumerate}
\item for $i\in\{0,1,\ldots, 95\}$
\item for $j\in\{0,1,\ldots, 95\}$
\item~\label{idsqq} if $RRot(A, i)\oplus RRot(A', j)==r\oplus r'$, return $RRot(A, i)\oplus r$, $i$ and $j$ respectively as $IDS$, $wt(q)$ and $wt(q')$.
\end{enumerate}
\item Given $r$, $r'$, $IDS$, $wt(q)$ and $wt(q')$, the adversary extracts $K$ as  $RRot((B\oplus Rot(IDS, wt(r))),wt(q))\oplus r$.
\item The adversary can use other metrics to filter possible wrong guesses, e.g., $B'$, $C$, $C'$, $D$ and $D'$. 
\item The adversary calculates the updated parameters as $IDS_{new}=Rot(IDS\oplus K, wt(r'))\oplus Rot(K\oplus wt(q'),wt(IDS))$ and $K_{new}=Rot(K, wt(r'))\oplus Rot(r'\oplus q',wt(K))$
\end{enumerate}	
Following above attack, the adversary extracts whole shared parameters between the tag and the reader with the complexity of eavesdropping two sessions of the protocol, blocking a message and doing polynomial computations. In this case also, since the adversary has whole secret parameters, applying any other attack will be trivial, e.g., reader/tag impersonation, desynchronization and traceability attacks.

\section{Conclusions}\label{conc} 
In this paper, we have shown that two recent proposed security protocols for IoT and RFID systems (i.e. \texttt{KSP} and \texttt{SOVNOKP} ) are vulnerable against secret disclosure attacks, were we presented attacks with the complexity of eavesdropping only two sessions of each protocol and negligible computational complexity.

\bibliographystyle{abbrv} 
\bibliography{wang-refrences} 
\end{document}